\documentclass{iaus}
\usepackage{graphicx}
\title[Disk-Loss and Disk Renewal Phases in Classical Be Stars II.] 
{Disk-Loss and Disk Renewal Phases in Classical Be Stars II.\\Detailed Analysis of Spectropolarimetric Data}

\author[Zachary H. Draper \& John P. Wisniewski]   
{Zachary H. Draper$^1$, John P. Wisniewski$^1$, Karen S. Bjorkman$^2$, Jon E. Bjorkman$^2$, Xavier Haubois$^3$, Alex C. Carciofi$^3$, Marilyn R. Meade$^4$}

\affiliation{$^1$Department of Astronomy, University of Washington, \\ email: {\tt zhd@u.washington.edu jwisnie@u.washington.edu} \\ [\affilskip]
$^2$Ritter Observatory, Department of Physics \& Astronomy, University of Toledo \\ [\affilskip]
$^3$IAG, Universidade de Sao Paulo\\ [\affilskip]
$^4$Space Astronomy Lab, University of Wisconsin-Madison}

\pubyear{2010}
\volume{272}  
\pagerange{119--120}
\jname{Active OB Stars: structure, evolution, mass-loss, and critical limits}
\editors{C. Neiner, G. Wade, G. Meynet, \& G. Peters eds.}
\begin{document}

\maketitle

\begin{abstract}
In \cite{wis_etal10}, paper I, we analyzed 15 years of spectroscopic and spectropolarimetric data from the Ritter and Pine Bluff Observatories of 2 Be stars, 60 Cygni and $\pi$ Aquarii, when a transition from Be to B star occurred. Here we anaylize the intrinsic polarization, where we observe loop-like structures caused by the rise and fall of the polarization Balmer Jump and continuum V-band polarization being mismatched temporaly with polarimetric outbursts. We also see polarization angle deviations from the mean, reported in paper I, which may be indicative of warps in the disk, blobs injected at an inclined orbit, or spiral density waves. We show our ongoing efforts to model time dependent behavior of the disk to constrain the phenomena, using 3D Monte Carlo radiative transfer codes.
\end{abstract}

\firstsection 
\section{Balmer Jump vs Continuum Polarization}

The time evolution of the intrinsic continuum V-band polarization (V-pol) of $\pi$ Aqr is shown in Figure 1.  We find evidence of clockwise loop-like structures (Figure 2) when comparing the evolution of the polarization across the Balmer Jump (BJ) vs V-pol, particularly during polarimetric outburst events (red in Figure 1). 60 Cyg also displays this behavior.

\begin{figure}[h]%
  \centering
  \parbox{0.32\textwidth}{\includegraphics[width=0.32\textwidth]{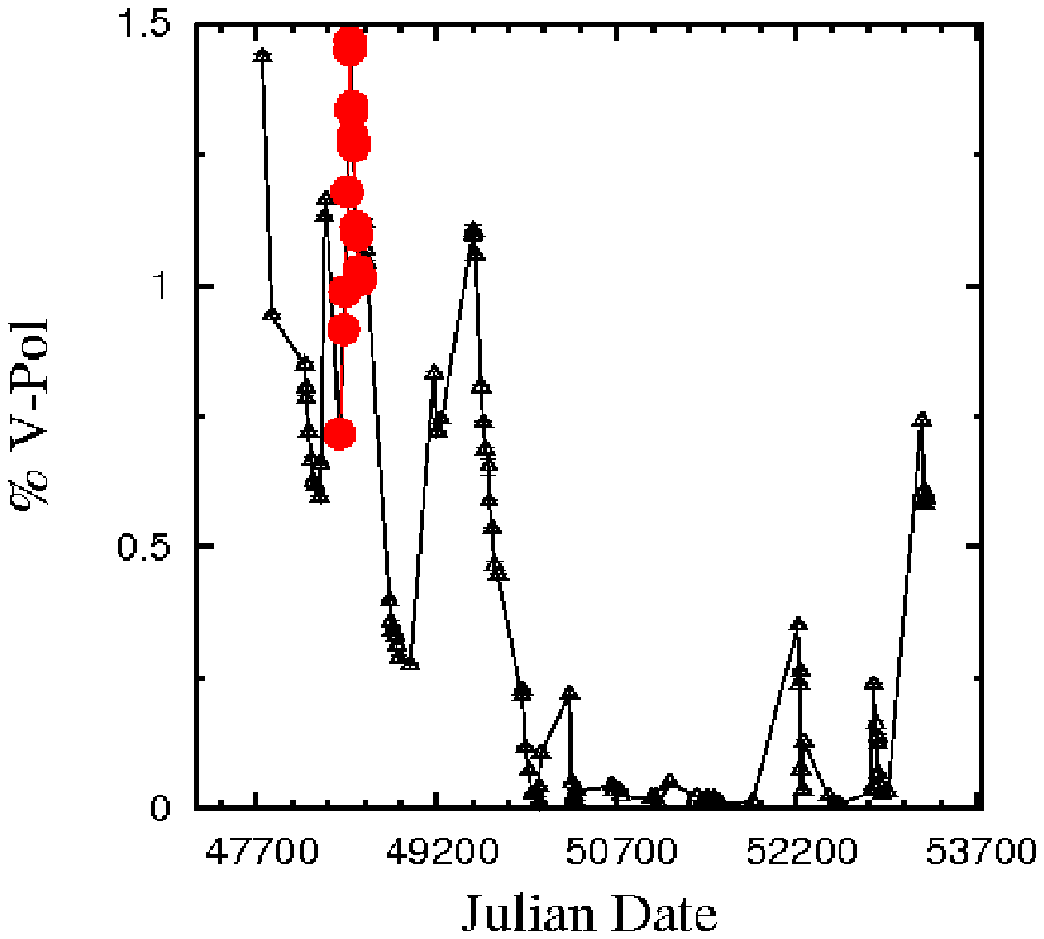}\caption{}}%
  \begin{minipage}{0.32\textwidth}%
     \includegraphics[width=1.0\textwidth]{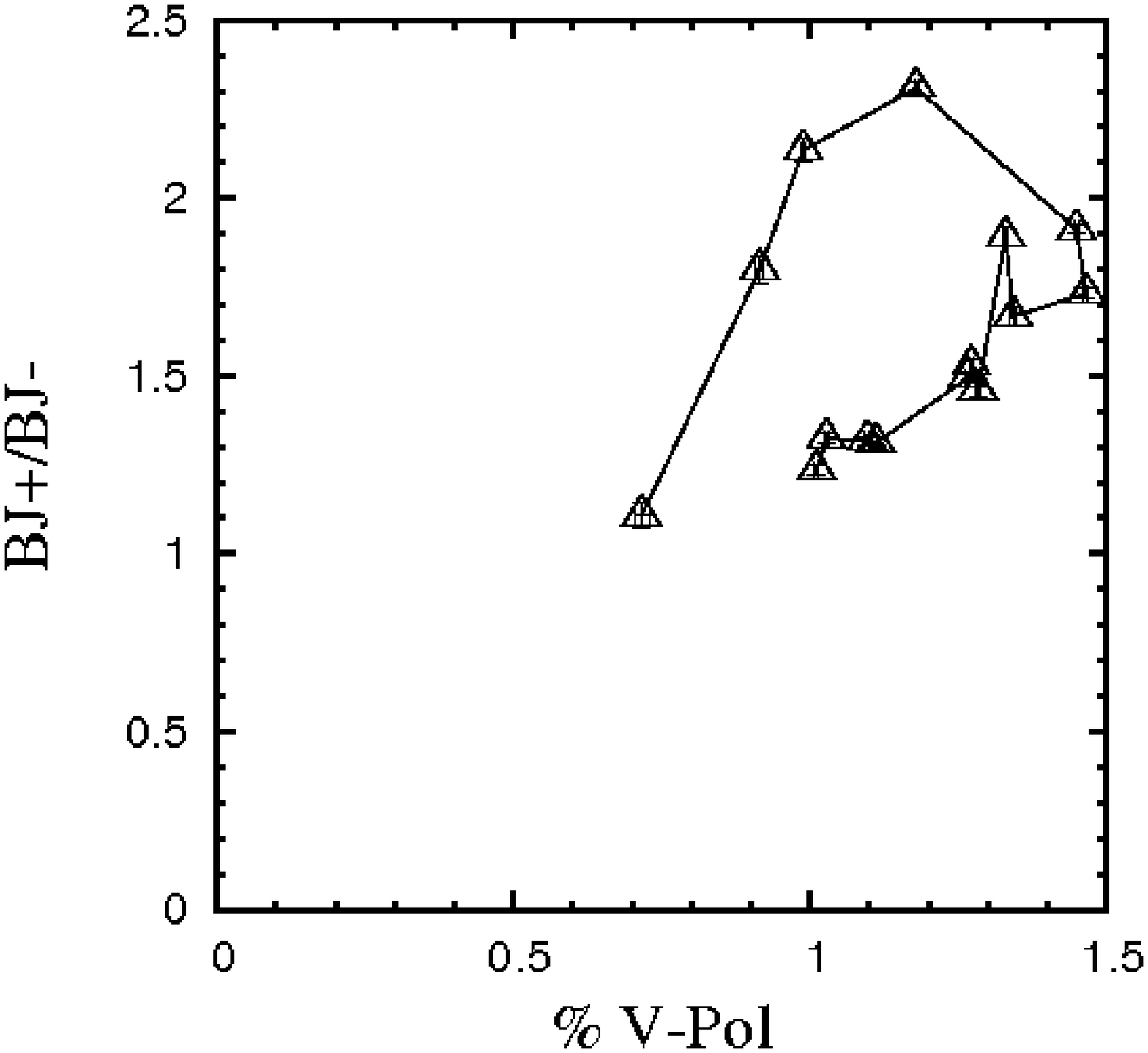}
     \caption{}
  \end{minipage}%
  \begin{minipage}{0.30\textwidth}%
     \includegraphics[width=1.0\textwidth]{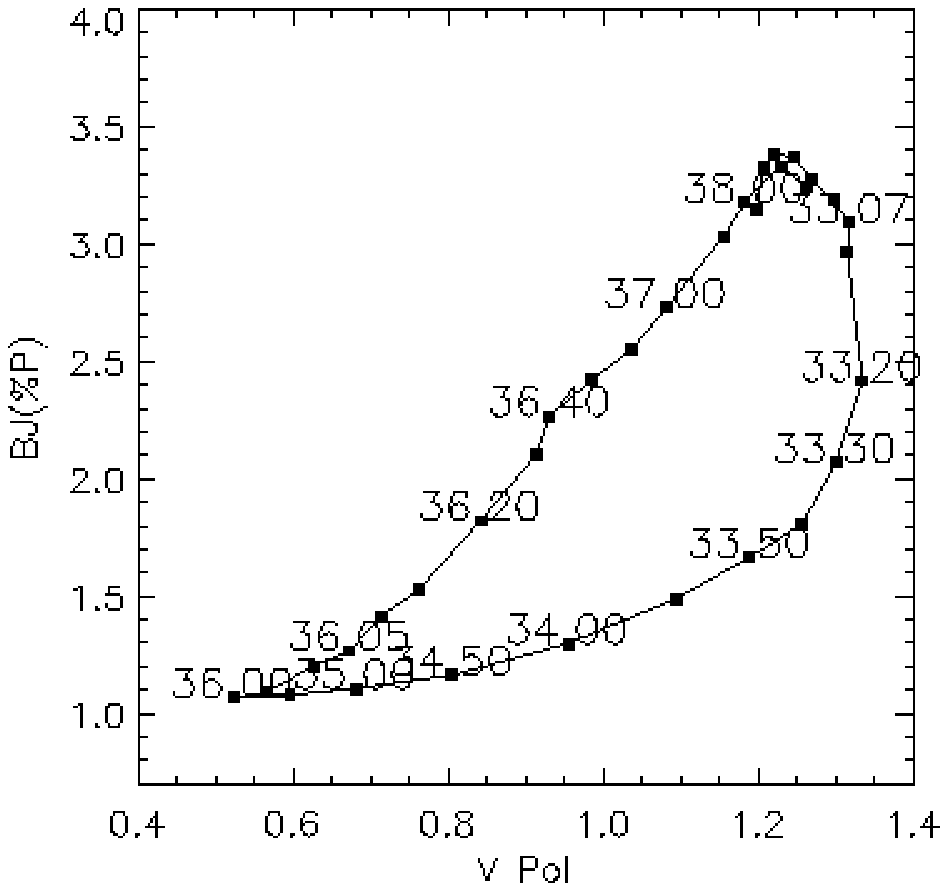}
     \caption{}
  \end{minipage}%
  \label{fig:1}%
\end{figure}

\section{Modeling Polarization Loops}

We use the non-LTE 3D Monte Carlo code developed by \cite{car_bjo06}, HDUST, to investigate the origin of the loop-like behavior of the BJ vs V-pol during polarimetric outbursts.  Interestingly, we find that the clockwise loop structures can be reproduced when the mass-loss from the central star which feeds the disk is turned on (6 to 12 o$^{'}$clock, Figure 3) then off (12 to 6 o$^{'}$clock, Figure 3).  We therefore suggest that this diagnostic can provide insight into the time dependence of the density of the innermost disk region.  Counter-clockwise loops are also observed in $\pi$ Aqr and 60 Cyg and will require further modeling to ascertain thier origin. We note that \cite{wit_etal06} found similar loop structures in CMD diagrams of Be stars. 

\section{PA Variations}

In \cite[paper I]{wis_etal10}, we detected variations in the PA of disks during polarimetric outbursts, and speculated that these variations could be indicative of warps, non-equatorial blob injections, or spiral density waves in the inner disk.  Figure 4 depicts deviations in the mean PA during polarimetric outbursts of $\pi$ Aqr.  As shown in Figure 5, we model a density enhancement on one side (orange) of the disk and a decrement on the opposing side (black) using HDUST. The result of computing the polarization from different viewing angles in a counter-clockwise motion are shown in Figure 6. This model suggests that asymeteric material being ejected from the star can potentially match observed deviations in the PA.

\begin{figure}[h]%
  \centering
  \parbox{0.33\textwidth}{\includegraphics[width=0.33\textwidth]{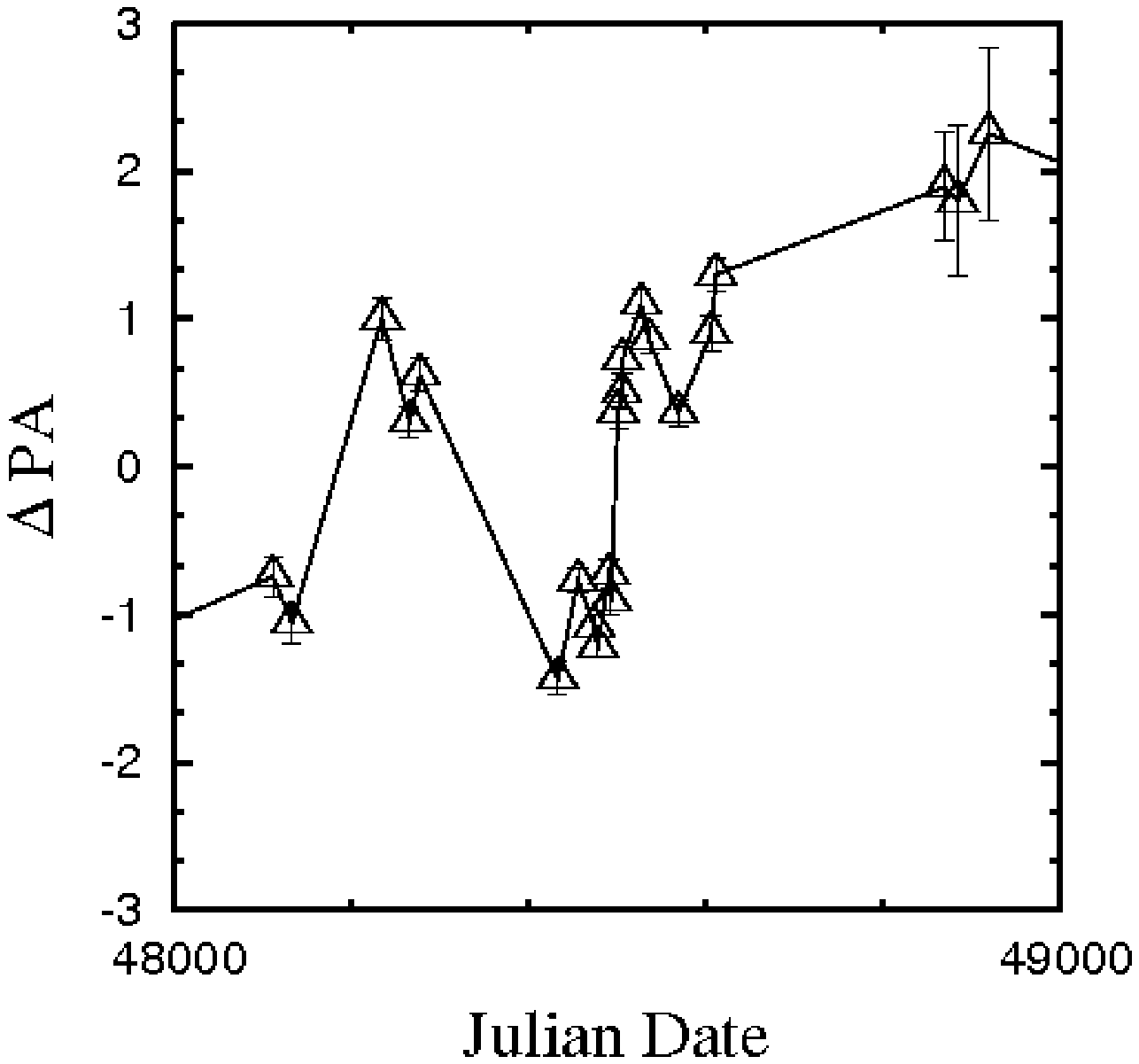}\caption{}}%
  \begin{minipage}{0.30\textwidth}%
     \includegraphics[width=1.0\textwidth]{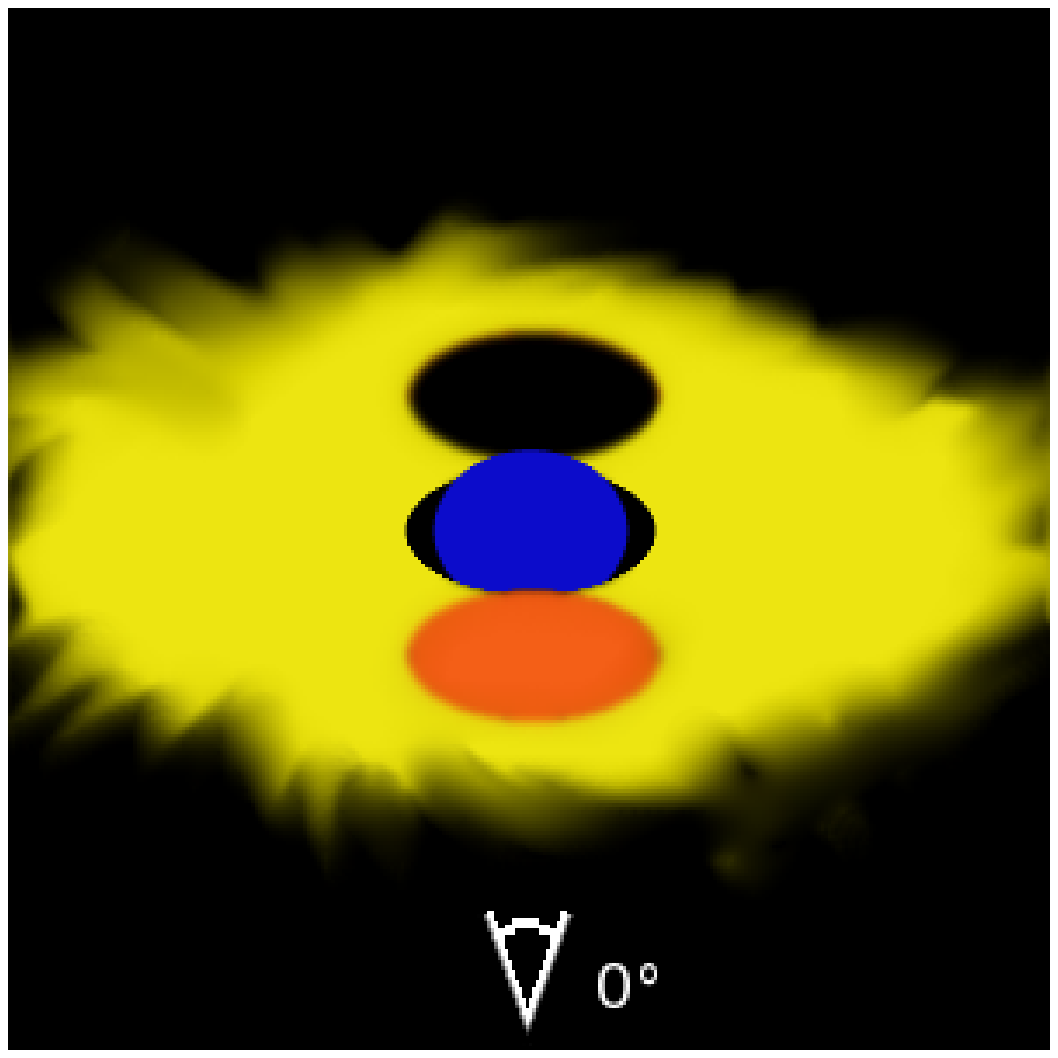}
     \caption{}
  \end{minipage}%
  \begin{minipage}{0.32\textwidth}%
     \includegraphics[width=1.0\textwidth]{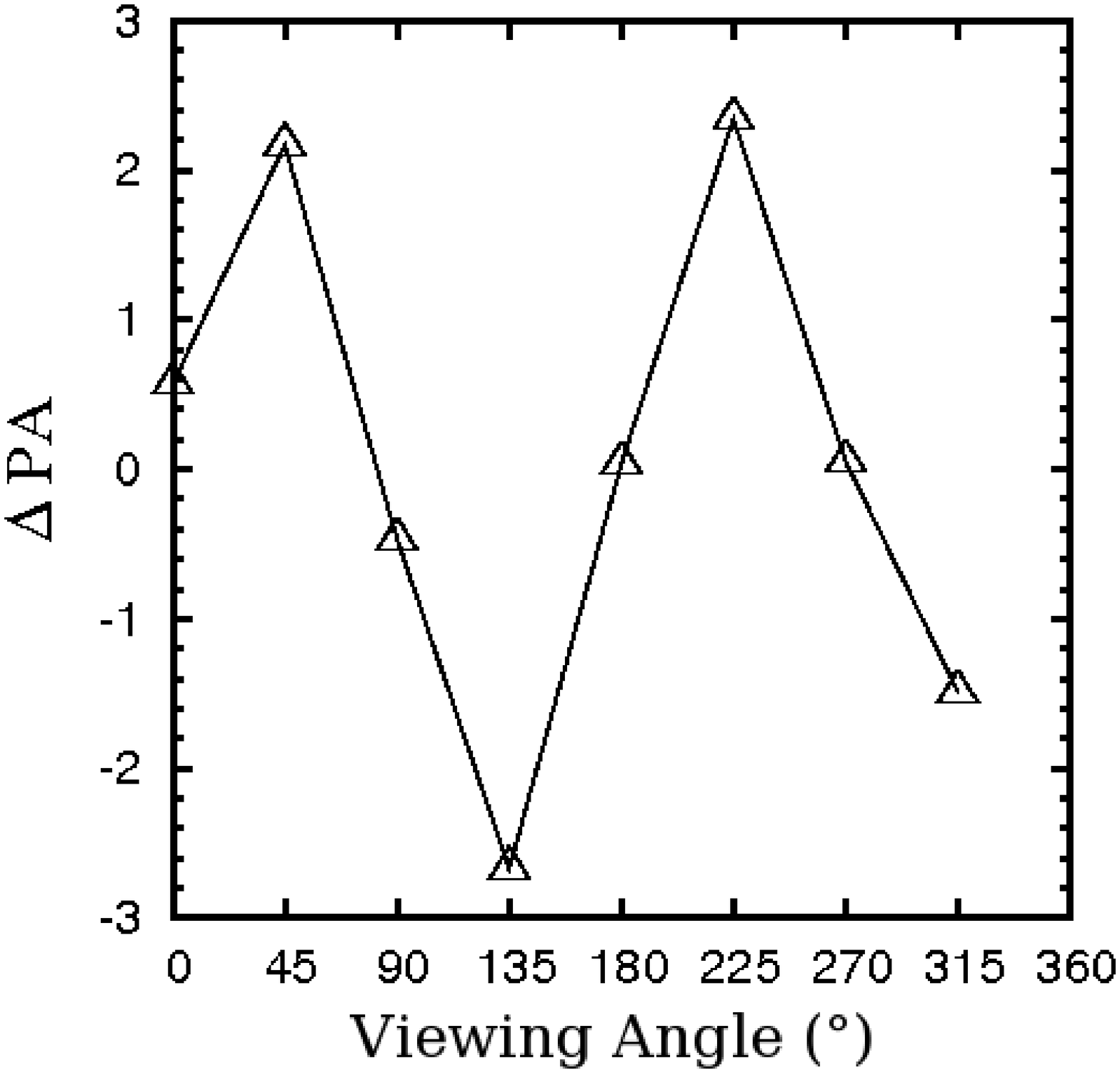}
     \caption{}
  \end{minipage}%
  \label{fig:2}%
\end{figure}

\end{document}